\documentclass[sigconf,nonacm]{acmart}

\usepackage{graphicx}
\usepackage{subcaption}
\usepackage{amsmath}
\usepackage{algorithm}
\usepackage[noend]{algpseudocode}
\usepackage{booktabs}            
\usepackage{adjustbox}           

\usepackage{dblfloatfix} 

\AtBeginDocument{%
  }

\newcommand{\ie}{\textit{i.e.,}\ }
\newcommand{\eg}{\textit{e.g.,}\ }

\begin{document}

\title[From Flows to Functions: Macroscopic Behavioral Fingerprinting of IoT Devices via Network Services]{From Flows to Functions: Macroscopic Behavioral Fingerprinting of IoT Devices via Network Services}

\author{Shayan Azizi}
\affiliation{%
	\institution{UNSW Sydney, Australia}
	\country{}
}

\author{Norihiro Okui}
\affiliation{%
	\institution{KDDI Research, Japan}
	\country{}
}

\author{Masataka Nakahara}
\affiliation{%
	\institution{KDDI Research, Japan}
	\country{}
}

\author{Ayumu Kubota}
\affiliation{%
	\institution{KDDI Research, Japan}
	\country{}
}

\author{Hassan Habibi Gharakheili}
\affiliation{%
	\institution{UNSW Sydney, Australia}
	\country{}
}

\renewcommand{\shortauthors}{S. Azizi, N. Okui, M. Nakahara, A. Kubota, and H. Habibi Gharakheili}

\begin{abstract}
Identifying devices such as cameras, printers, voice assistants, or health monitoring sensors, collectively known as the Internet of Things (IoT), within a network is a critical operational task, particularly to manage the cyber risks they introduce. While behavioral fingerprinting based on network traffic analysis has shown promise, most existing approaches rely on machine learning (ML) techniques applied to fine-grained features of short-lived traffic units (packets and/or flows). These methods tend to be computationally expensive, sensitive to traffic measurement errors, and often produce opaque inferences. In this paper, we propose a macroscopic, lightweight, and explainable alternative to behavioral fingerprinting focusing on the network services (\eg TCP/80, UDP/53) that IoT devices use to perform their intended functions over extended periods. Our contributions are threefold. 
(1) We demonstrate that IoT devices exhibit stable and distinguishable patterns in their use of network services over a period of time. We formalize the notion of service-level fingerprints and derive a generalized method to represent network behaviors using a configurable granularity parameter. 
(2) We develop a procedure to extract service-level fingerprints, apply it to traffic from 13 consumer IoT device types in a lab testbed, and evaluate the resulting representations in terms of their convergence and recurrence properties. 
(3) We validate the efficacy of service-level fingerprints for device identification in closed-set and open-set scenarios. Our findings are based on a large dataset comprising about 10 million IPFIX flow records collected over a 1.5-year period.

\end{abstract}



\maketitle

\section{Introduction}
The rapid growth of IoT devices has introduced cybersecurity challenges \cite{Gartner:IoTSec} in network asset management. These devices often lack visibility, rely on proprietary or undocumented protocols, and connect to external cloud services beyond local control. Consequently, identifying and classifying IoT devices has become essential for network operators who want to enforce security policies \cite{Feamster:Blog17, techtarget:CSAM}, manage traffic, and detect rogue devices.

Existing research in IoT traffic classification has largely focused on microscopic approaches that use machine learning to classify devices based on short-lived packet-level features \cite{Sivanathan:TMC19, Miettinen:ICDCS17} or flow-level metadata \cite{Pashamokhtari:LCN21}. Although effective under controlled conditions, these methods often require fine-grained traffic data (\eg packet headers, inter-arrival times, or payloads), involve computationally intensive feature extraction, and are sensitive to telemetry errors and noise. Moreover, they frequently encounter high-class overlap at the packet or flow level. Additionally, their inference processes are typically opaque, which limits interpretability and hinders practical deployment in real-time network environments.

From an operational perspective, the goal of IoT fingerprinting extends beyond labeling individual flows or packets; it is to map connected devices and understand their macroscopic behavior over an extended period. Network administrators are typically more concerned with identifying \textit{what} a device is (make/model or category) and \textit{how} it behaves on the network, rather than analyzing transient properties of isolated traffic. This requires behavioral fingerprinting techniques that capture the functional role of a device and long-term communication patterns, rather than just immediate traffic characteristics.

In this work, we propose a lightweight alternative to traditional flow-level classification by focusing on the network services that IoT devices access (\eg TCP/80, UDP/53). These service-level interactions reflect how a device functions. and are inherently more stable and interpretable than individual flows. We refer to the aggregate pattern of these interactions as a device's service-level fingerprint, a compact yet expressive representation of its profile. 

Importantly, our approach aligns with the Manufacturer Usage Description (MUD) standard proposed by the IETF \cite{RFC8520}, which defines a declarative model for expected IoT device behavior on a network. MUD profiles, specified by manufacturers, outline permitted endpoints, protocols, and transport-layer ports. By capturing behavioral patterns at a similar level of abstraction, service-level fingerprinting supports device classification and also offers a practical basis for validating MUD profiles, detecting behavioral deviations, and generating policy templates in MUD-like formats.    

This paper makes the following specific contributions. Our \textbf{first} contribution (\S\ref{sec:contribution1}) demonstrates that IoT devices exhibit distinct and consistent usage patterns of network services over time. Based on our observations of traffic patterns from two representative device types, we formalize three representations (\ie Service List, Service Prevalence, and Generalized) that capture different aspects of service usage.
Our \textbf{second} contribution (\S\ref{sec:contribution2}) develops a method to extract service-level fingerprints from flow data and applies it to traffic collected from 13 IoT device types on a lab testbed. We evaluate the fidelity of the extracted fingerprints using quantitative metrics of convergence and recurrence.
Our \textbf{third} contribution (\S\ref{sec:contribution3}) validates the efficacy of the extracted fingerprints to the task of classifying device types in closed-set (13 seen classes) and open-set scenarios (13 seen and 22 unseen classes), demonstrating high precision and recall. Our findings are based on a dataset of 10 million IPFIX flow records, spanning 35 device classes and collected over 1.5 years.

\section{Our IoT Traffic Dataset}
\textbf{Data Collection Setup:} We use IPFIX flow records of IoT devices provided by an ISP, collected from a lab testbed over a 1.5-year period, spanning mid-2019 to the end of 2020. For this study, we start with 13 unique types of IoT devices (make-and-model classes) that remained active throughout the data collection period.  In addition, we incorporate 22 more device types that were introduced in 2020. Each IPFIX record contains header-derived identifiers and fields, such as MAC / IP addresses, transport layer protocols, and port numbers. While the records also include flow-level metadata, such as counters and timers, these fields are not utilized in our analysis. This study was conducted in accordance with the ethical guidelines of our institute, and all necessary approvals were obtained.

\textbf{Data Preparation:} Based on our analysis we noticed that a logical communication flow may be exported as multiple IPFIX records. This can occur due to packet drops or delays, which may trigger premature flow timeouts or because of previously delayed packets that arrive after a session has been forcefully terminated. We removed redundant flows by tracking the temporal proximity of exported flows using their identifiers. This data preparation is crucial for this study, which focuses on the temporal usage patterns of network services, since redundant flows could distort these statistics. Also, we discarded all local and IPv6 traffic. The resulting dataset contains approximately 10 million IPFIX records used in this analysis.

\section{Network Service Usage Behaviors}
\label{sec:contribution1}
In this section, we motivate our research by analyzing the behavior of two representative IoT devices over 24 weeks. We show that their network service usage compositions produce identifiable fingerprints. We then discuss methods to represent these service-level fingerprints quantitatively.

\vspace{-2mm}
\subsection{Service Usage Diversities}
\begin{figure*}[ht]
  \centering

  \begin{subfigure}[t]{0.48\linewidth}
    \centering
    \includegraphics[width=\linewidth]{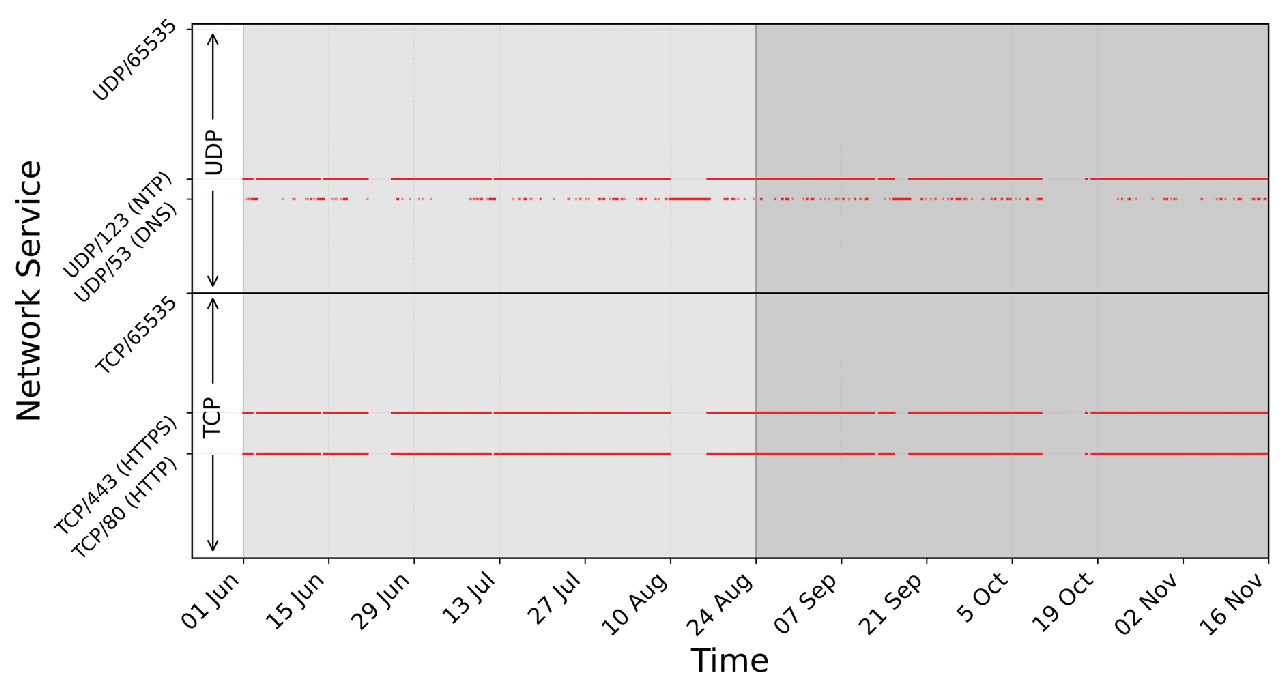}
    \caption{Temporal service usage in traffic of Amazon Echo.}
    \label{fig:serviceTraceAmazonEcho}
  \end{subfigure}
  \hfill
  \begin{subfigure}[t]{0.48\linewidth}
    \centering
    \includegraphics[width=\linewidth]{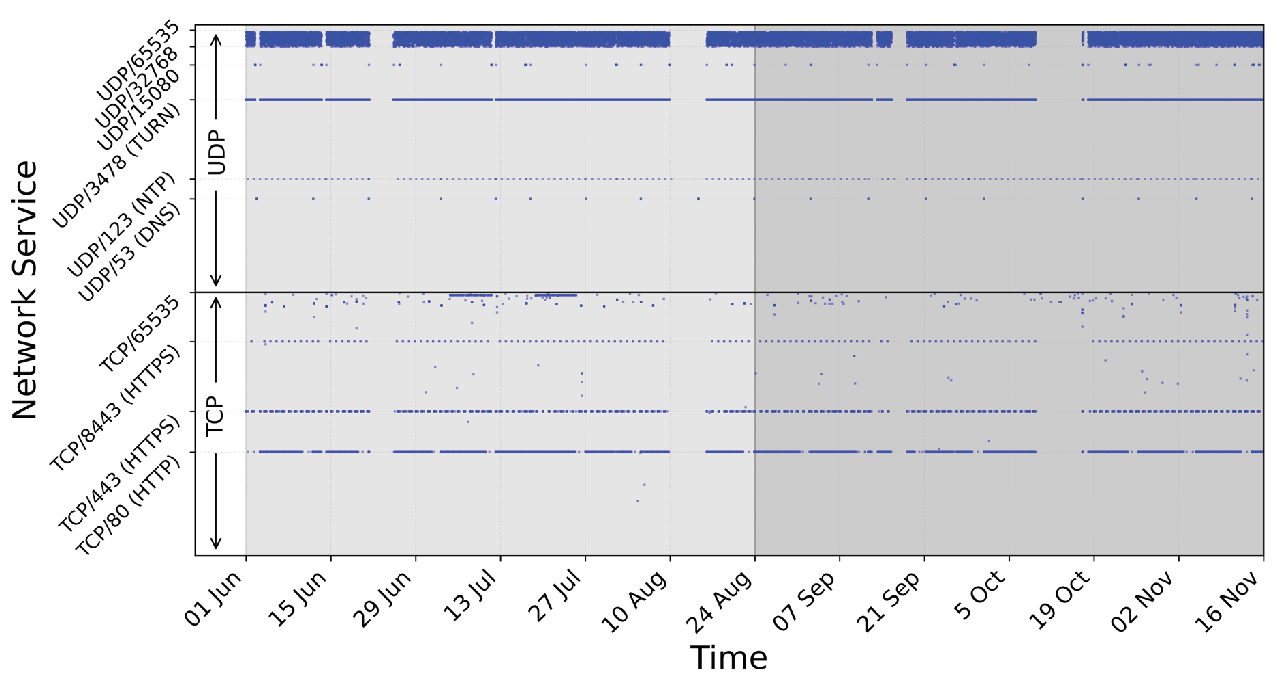}
    \caption{Temporal service usage in traffic of QWatch Cam.}
    \label{fig:serviceTraceQWatchIPCamera}
  \end{subfigure}

  \medskip

  \begin{subfigure}[t]{0.48\linewidth}
    \centering
    \includegraphics[width=\linewidth]{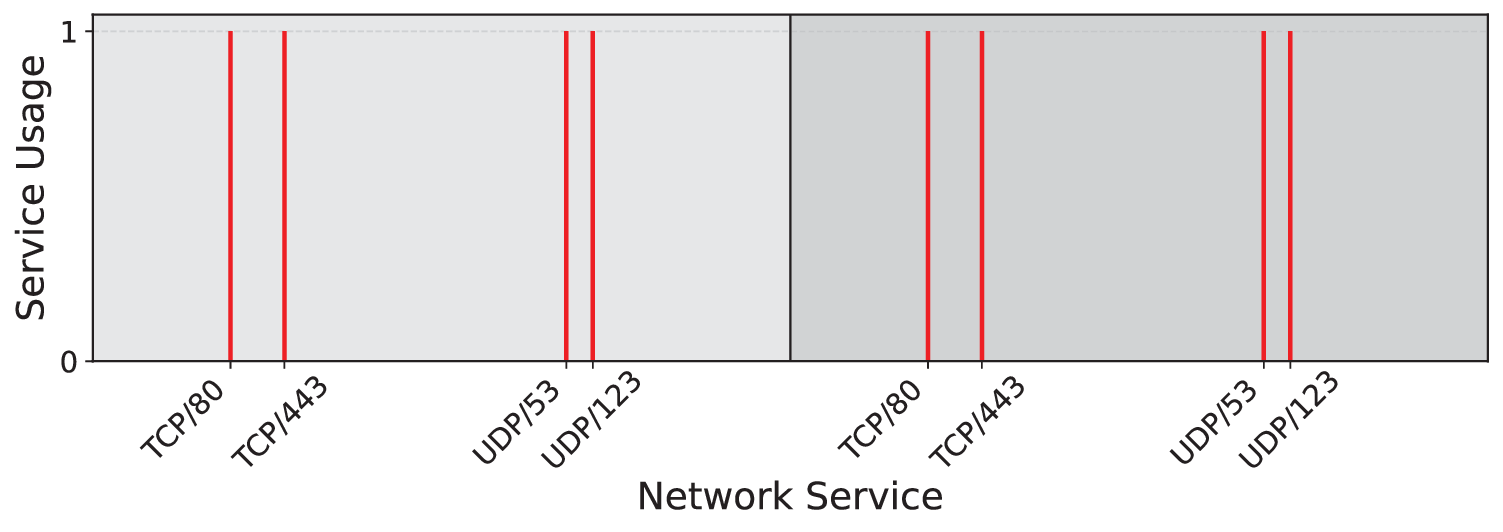}
    \caption{SL representation for Amazon Echo.}
    \label{fig:SLrep4AmazonEcho}
  \end{subfigure}
  \hfill
  \begin{subfigure}[t]{0.48\linewidth}
    \centering\includegraphics[width=\linewidth]{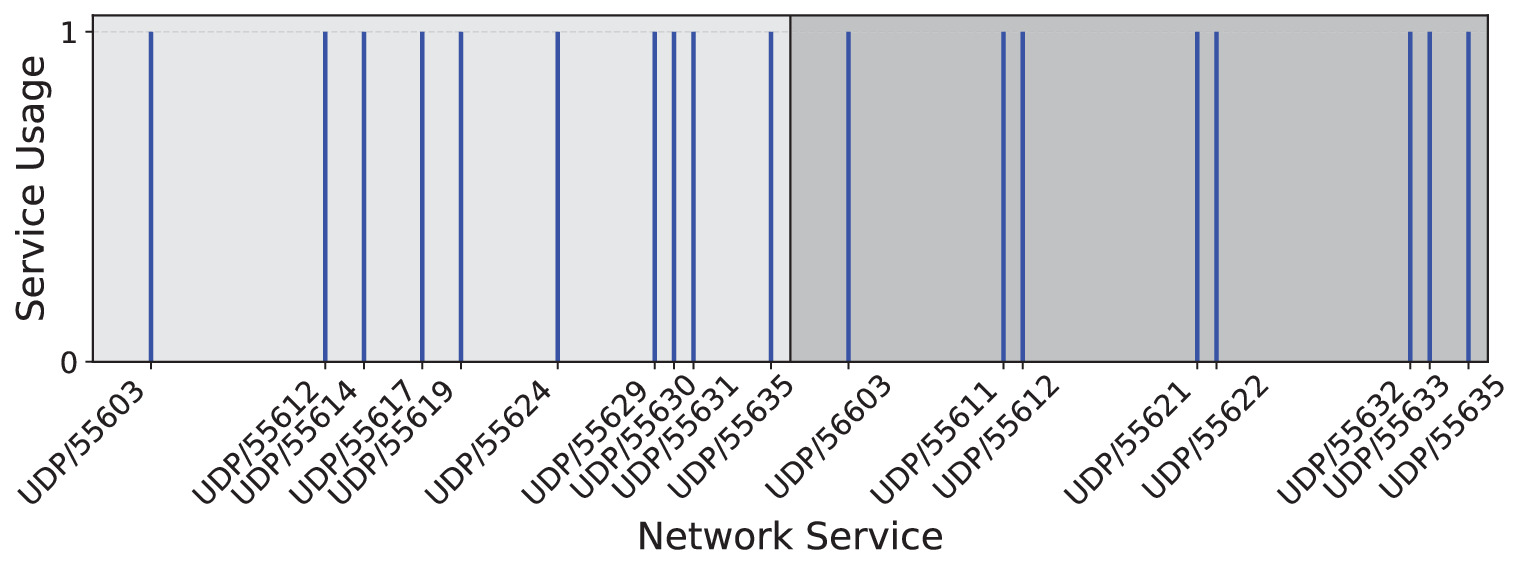}
    \caption{SL representations (cropped) for QWatch Cam.}
    \label{fig:SLrep4QWatchIPCamera}
  \end{subfigure}

  \medskip

  \begin{subfigure}[t]{0.48\linewidth}
    \centering
    \includegraphics[width=\linewidth]{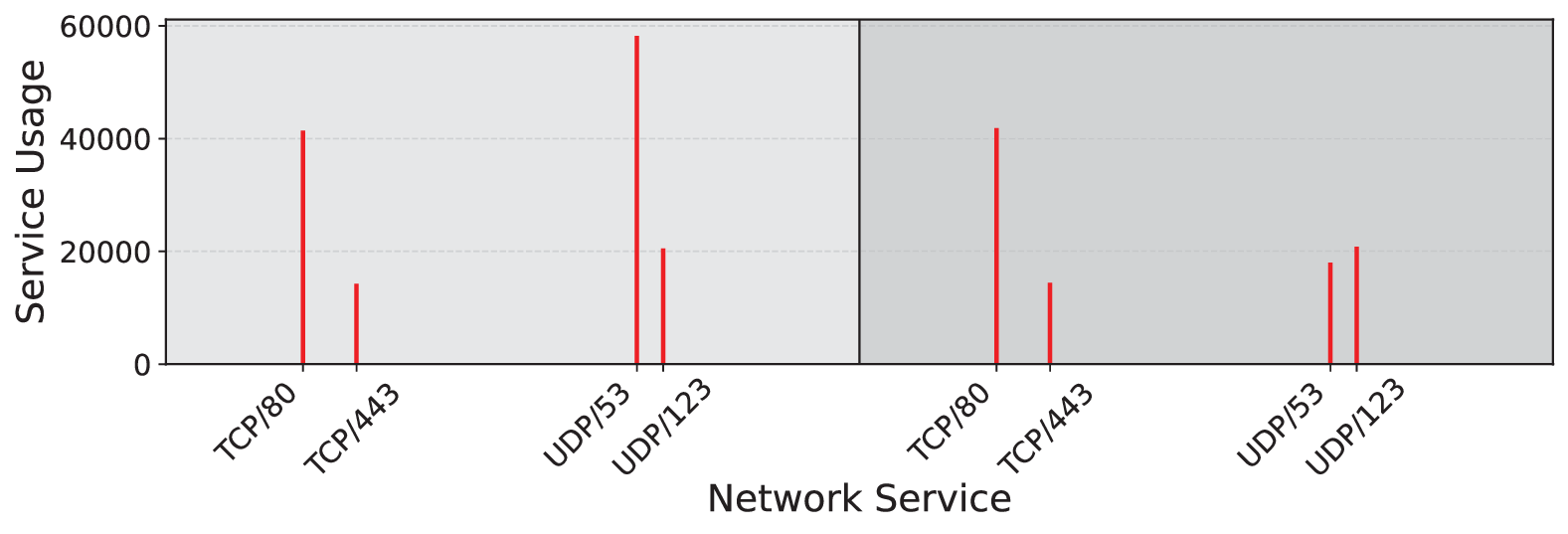}
    \caption{SP representation for Amazon Echo.}
    \label{fig:SPrep4AmazonEcho}
  \end{subfigure}
  \hfill
  \begin{subfigure}[t]{0.48\linewidth}
    \centering
    \includegraphics[width=\linewidth]{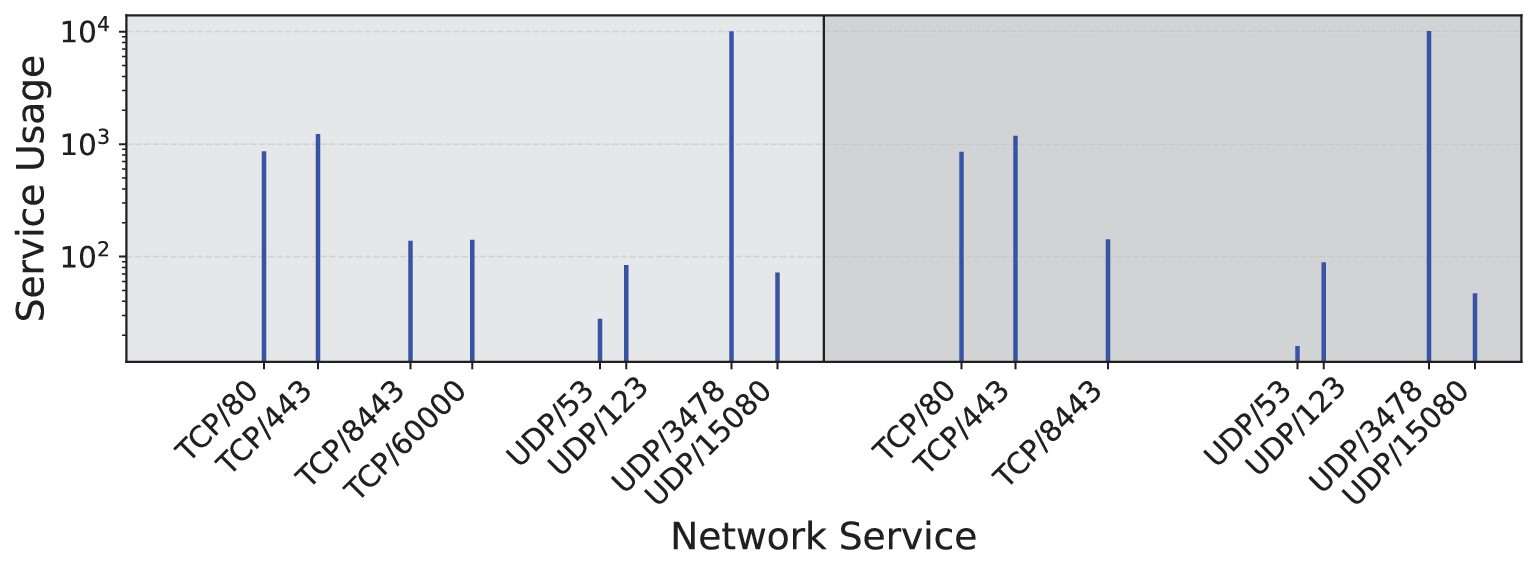}
    \caption{SP representation for QWatch Cam.}
    \label{fig:SPrep4QWatchIPCamera}
  \end{subfigure}

  \medskip

  \begin{subfigure}[t]{0.48\linewidth}
    \centering
    \includegraphics[width=\linewidth]{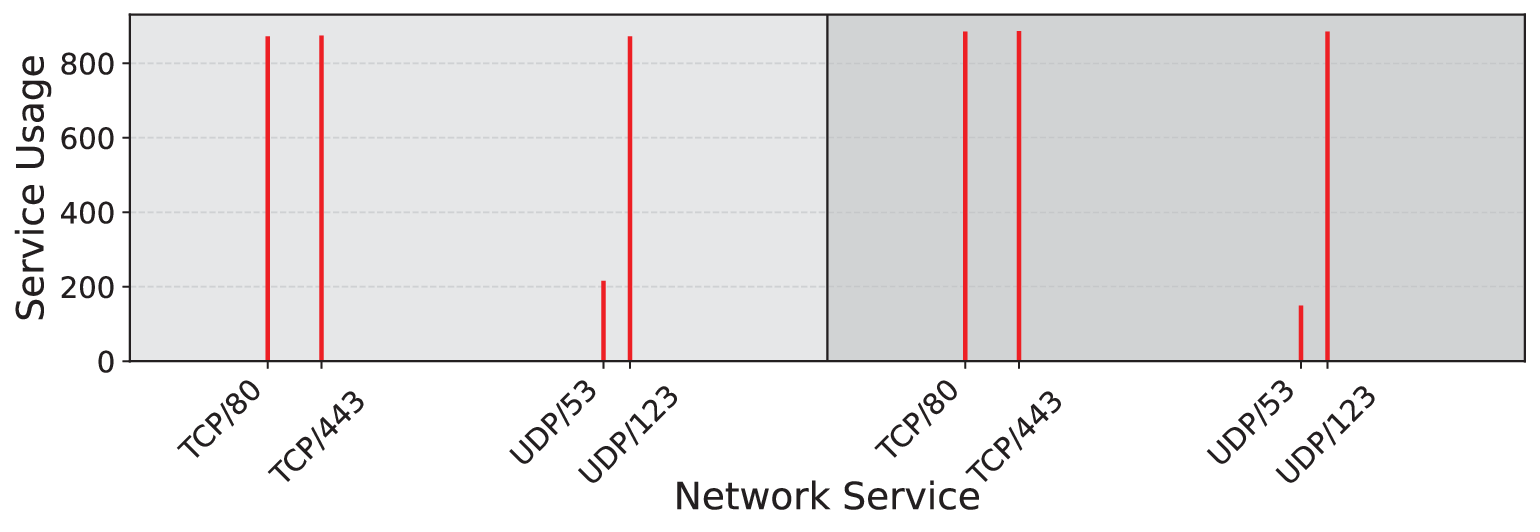}
    \caption{G represent. with $g=1024$ for Amazon Echo.}
    \label{fig:Grep4AmazonEcho}
  \end{subfigure}
  \hfill
  \begin{subfigure}[t]{0.48\linewidth}
    \centering
    \includegraphics[width=\linewidth]{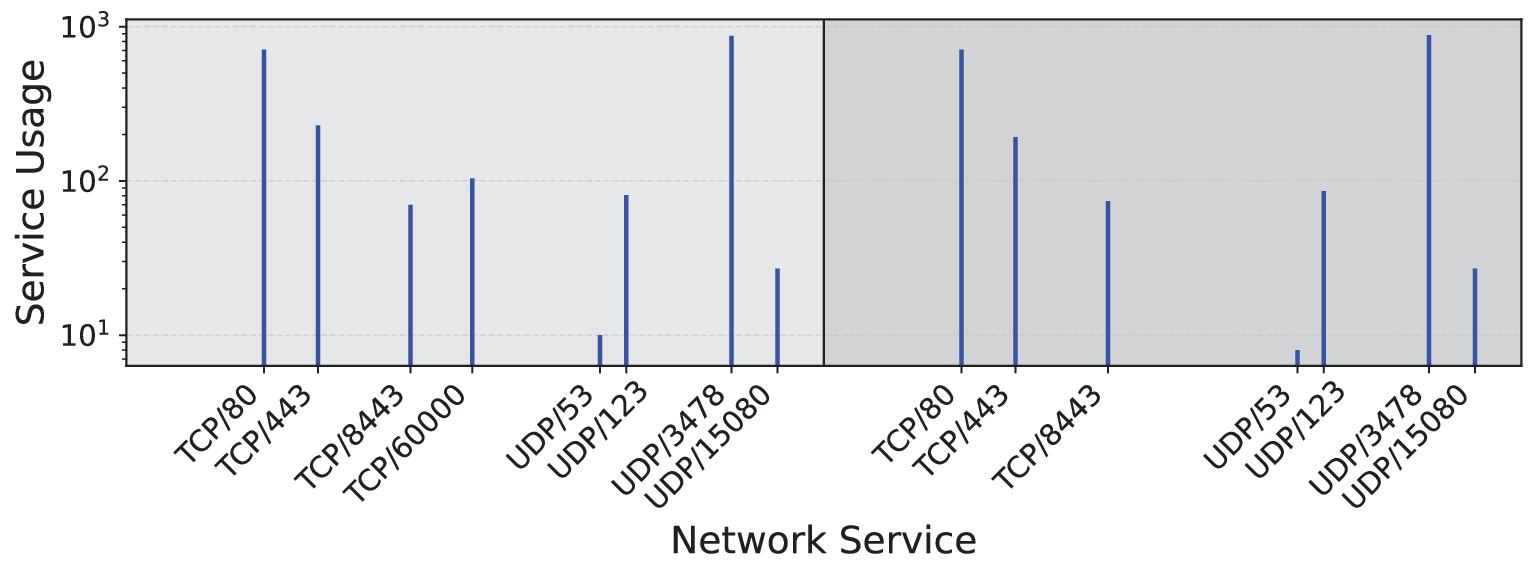}
    \caption{G represent. with $g=1024$ for QWatch Cam.}
    \label{fig:Grep4QWatchCamera}
  \end{subfigure}
  \vspace{-3mm}  
  \caption{IoT devices exhibit distinct patterns (a,b) in their usage of network services. While SL representations may yield stable fingerprints for some devices (c), they are not helpful for some other devices that communicate with a large range of dynamic or Ephemeral destination ports (d). SP representation can stabilize the mentioned behavior (f), but could be too sensitive to variabilities in network usage (e). The G representations, when used with an appropriate choice of \textit{granularity level}, are capable of addressing the mentioned shortfalls (g, h).}  \label{fig:visualisingFingerprintsAndRepresentations}
\end{figure*}

Each IoT device relies on a set of network services to function. 
Figures \ref{fig:serviceTraceAmazonEcho} and  \ref{fig:serviceTraceQWatchIPCamera} illustrate the service usage traces of the Amazon Echo and the I-O DATA Qwatch IP Camera, respectively. These traces span 24 weeks from June 1st to November 16th, 2019. Each data point represents an IPFIX flow, with bolder areas indicating higher usage of the service.

It can be seen that Amazon Echo (Fig. \ref{fig:serviceTraceAmazonEcho}) exhibits a relatively simple network activity, relying solely on four network services: TCP/80 (likely HTTP), TCP/443 (likely HTTPS), UDP/53 (likely DNS) and UDP/123 (likely NTP). Among these, the use of TCP/80, TCP/443, and UDP/123 is relatively consistent, but UDP/53 is used intermittently. 
In contrast, the behavior of the QWatch Camera (Fig. \ref{fig:serviceTraceQWatchIPCamera}) is more complex. Like the Amazon Echo, it uses TCP/80 and TCP/443, but it also consistently communicates over TCP/8443, a common alternative port for HTTPS. 
Additionally, the camera utilizes various sporadic TCP services, primarily within the high port range of TCP/37000---TCP/65535. On the UDP side, the device shares similarities with the Amazon Echo by using UDP/123 and occasionally UDP/53. However, it also relies on UDP/3478 (likely TURN) and UDP/15080, which is likely a proprietary service. Furthermore, it frequently accesses a wide range of sporadic UDP ports between 32700 and 61000. The QWatch Camera's extensive use of TURN, combined with its high port-range activity, suggests involvement in NAT traversal, peer-to-peer communication, and possibly Real-time Transport Protocol (RTP) traffic, typical behavior of an IP camera. Based on these high-level observations, our next objective is to formalize a method for representing service usage patterns that can serve as behavioral fingerprints for device identification.

\subsection{Representation of Service Usage Compositions}
\label{sec:summarization}
We demonstrated the potential of service usage compositions for fingerprinting the network behavior of IoT devices. However, a key question remains unanswered: \textit{how can an IoT device's service usage composition be quantitatively represented?} In this section, we develop a general framework for representing service usage compositions, which produces fingerprints at different levels of precision based on a parameter we define as the \textit{granularity level}, denoted by $g$. As we will demonstrate throughout the paper, the choice of $g$ plays a crucial role in the effectiveness of the fingerprint.

To start, consider a window of observed flows from an IoT device over a given period. A natural way to represent the device's service usage composition is to capture the list of distinct network services it utilizes, regardless of their prevalence---a method we refer to as the Service List (SL) representation. 
Let there be a one-to-one mapping $\pi$ from all possible UDP and TCP network services to indices in the range $\{1,2,\ldots,2\times2^{16}\}$.
Given a window $W$ of observed network services $s_1, s_2, \ldots, s_n$, we define the SL representation, denoted by $\mathbb{\mathrm{r}}_{SL}(W)$, as a binary vector of length $2\times2^{16}$.  Each element $i$ of this vector is defined as follows:

\begin{equation}
\label{Eq: Service Set Fingerprint}
r_{SL}^{(i)}(W) =
\begin{cases}
  1 & \text{if } \pi(s_j) = i \; for \; some \; j\in [n], \\
  0  & \text{otherwise}.
\end{cases}
\end{equation}

As the definition implies, SL representations can be \textit{too coarse-grained}, which limits their effectiveness for fingerprinting. To illustrate this limitation, we divide the observation period in Fig. \ref{fig:visualisingFingerprintsAndRepresentations} into two consecutive windows (indicated by lighter and darker background shades). The corresponding SL representations for both windows are shown in Fig. \ref{fig:SLrep4AmazonEcho}  for the Amazon Echo and Fig. \ref{fig:SLrep4QWatchIPCamera} for the QWatch Camera. 
For better readability, the SL representation of the QWatch Camera is cropped to the UDP port range 55603---55635. In the case of the Amazon Echo, the SL representations for both windows are identical, reflecting consistent use of just four services: TCP/80, TCP/443, UDP/53 and UDP/123. This suggests a stable and compact service usage pattern over time. In contrast, the QWatch Camera shows a significant difference between the SL representations of the two windows. The sporadic services observed in the first window largely differ from those seen in the second, resulting in minimal overlap. This highly variable behavior makes it challenging to define a stable fingerprint using SL representation. Instead of converging to a compact set of services within a reasonable time frame, the service list continues to grow (bounded only by the full TCP/UDP port range), making SL an impractical choice for fingerprinting dynamic devices.

To quantitatively assess the similarity between two representation vectors, we use cosine similarity in this paper. For example, the cosine similarity between the SL representations of consecutive windows for the Amazon Echo and the QWatch Camera is $1$ and $0.28$, respectively. Widely used in fields such as Natural Language Processing (NLP), cosine similarity is invariant to vector magnitude and depends solely on the angle (orientation) between vectors. This property ensures that our similarity measure is not affected by the overall volume of traffic used to construct the representation. As a result, it enables meaningful comparisons between windows of varying durations. Furthermore, even when comparing windows of equal length, the level of activity from a device may vary significantly for reasons such as changes in user interactions. Cosine similarity remains robust in such cases, offering a consistent metric for evaluating representation similarity regardless of traffic intensity or window duration.

Given the consistent use of the QWatch Camera's core services (TCP/80, TCP/443, and UDP/3478), an effective way to filter out the sporadic services is to count the frequency with which each service appears within a given window. 
Following the same notation as before, we define the Service Prevalence (SP) representation, denoted by $\mathbb{\mathrm{r}}_{SP}$, as a vector of length $2\times2^{16}$. Each element $i$ is defined as follows:

\begin{equation}
\label{Eq: Service Prevalence Fingerprints}
    r_{SP}^{(i)}(W) = \left|\left\{j|\pi(s_j)=i, j\in [n]\right\}\right|.
\end{equation}

where, $r_{SP}^{(i)}(W)$ counts the number of times the service corresponding to index $i$ (as determined by the mapping $\pi$) appears in window $W$.  

As shown in Fig.~\ref{fig:SPrep4QWatchIPCamera}, the SP representation effectively suppresses the influence of sporadic services and captures the core fingerprint of the device. This fingerprint remains consistent across the two windows, except TCP/60000, which is absent in the second window. With this refined representation, the cosine similarity between the two representations increases dramatically from $0.28$ (in the SL representation) to $0.9998$, highlighting the stability in the usage of core services. However, in some cases, the SP representation can be \textit{too precise} or \textit{overly granular}, reducing its robustness to transient fluctuations in traffic. This issue is illustrated by the SP representations of Amazon Echo across two windows in Fig. \ref{fig:SPrep4AmazonEcho}. While the usage of TCP/80, TCP/443 and UDP/123 remains nearly identical, the prevalence of UDP/53 is almost three times higher in the first window compared to the second. This discrepancy is largely due to a temporary surge in DNS queries between August 10th and 14th (possibly due to temporary shifts in cloud endpoints after an update), which skews the SP vector and gives an impression of DNS-heavy behavior in the first window. However, as shown in the ground-truth trace in Fig. \ref{fig:serviceTraceAmazonEcho}, the overall UDP/53 activity seems quite similar across both windows. This suggests that SP, while valuable for emphasizing consistent patterns, may be too sensitive to short-term traffic spikes that do not reflect long-term behavior. 

Based on our observations, an effective representation of service usage composition should strike a balance between responsiveness to the relative importance of different network services and robustness to short-term variability. To achieve this, we propose a Generalized (G) representation, which captures the consistency of the service usage rather than relying solely on total count (as in SP) or binary presence (as in SL). The granularity of this representation is controlled by a parameter called the \textit{granularity level} ($g$). To calculate the G representation, denoted by $\mathbb{\mathrm{r}}_{G}$, for a given window $W$ of network flows, we divide $W$ into $g$ consecutive, equal-length sub-windows $w_1, w_2,\ldots,w_g$. For each service, we then count the number of sub-windows in which that service appears at least once. More precisely, we define the Generalized representation as follows:

\begin{equation}
\label{Eq: Smoothed Service Prevalence}
\mathbb{\mathrm{r}}_{G}(W;g) = \sum_{k=1}^{g} \mathbb{\mathrm{r}}_{SL}(w_k).
\end{equation}

This formulation allows for the interpretation of the Generalized representation as a smoothed service prevalence measure, quantifying how consistently each service appears across the sub-windows. 
Note the following special cases. When $g=1$,  the entire window is treated as a single block, reducing the Generalized representation to the SL representation: $\mathbb{\mathrm{r}}_{G}(W;g=1) = \mathbb{\mathrm{r}}_{SL}(W)$. 
As $g \to \infty$, each sub-window becomes so narrow that it captures only one service instance. In this limit, the Generalized representation approximates the Service Prevalence (SP) representation:
$ \mathbb{\mathrm{r}}_{G}(W;\infty) = \mathbb{\mathrm{r}}_{SP}(W)$.

The Generalized (G)  representations for Amazon Echo and QWatch Camera are shown in Fig. \ref{fig:Grep4AmazonEcho} and Fig. \ref{fig:Grep4QWatchCamera}, respectively. For Amazon Echo, the three main services (TCP/80, TCP/443 and UDP/123) appear with nearly equal consistency throughout each window when the granularity level is set to $g = 1024$. In contrast, UDP/53, is projected as a less persistent service, aligning well with its intermittent usage pattern shown in Fig. \ref{fig:serviceTraceAmazonEcho}. Overall, the G representation remains stable across the two time windows, yielding a high cosine similarity of $0.999$. 

Similarly, for the QWatch Camera, the G representations corresponding to the two windows are highly consistent, with a cosine similarity of $0.987$. An interesting comparison between  Fig. \ref{fig:SPrep4QWatchIPCamera} (SP representation) and Fig. \ref{fig:Grep4QWatchCamera} (G representation) reveals a subtle shift in interpretation: while the SP representation indicates a slightly higher prevalence of TCP/443 services over TCP/80, the G representation highlights significantly greater consistency in the use of TCP/80. This observation better reflects the frequent interruptions in TCP/443 usage as seen in Fig. \ref{fig:serviceTraceQWatchIPCamera}. This contrast illustrates how G representation captures the temporal stability of service usage more effectively than raw frequency counts, offering a robust fingerprint of device behavior.

\section{Exporting and Assessing Service Level Fingerprints}
\label{sec:contribution2}
In the previous section, we demonstrated how IoT devices differ in their use of network services and developed a generalized representation method to capture their service-level fingerprints. In this section, we develop a methodology to export fingerprints of IoT devices from their network traffic flow records. We also propose metrics to evaluate the impact of different representations on the convergence and recurrence quality of the resulting fingerprints.

\subsection{Fingerprint Exporter Design}
\label{subsec:exporter}
Our previous discussions mainly focused on \textit{how to quantitatively represent} service-level behaviors effectively. Another important requirement for fingerprinting is addressing the following question: \textit{How much of a device’s observed traffic is sufficient to export a reliable service-level fingerprint?} 

The temporal patterns of service utilization vary significantly across devices (\eg compare Fig. \ref{fig:serviceTraceAmazonEcho} and Fig. \ref{fig:serviceTraceQWatchIPCamera}), with some services used consistently and others only intermittently. Hence, a search procedure is needed to identify the appropriate observation window that captures the full spectrum and seasonality of each device's network activity. The device fingerprint will be exported based on the outcome of this search, ensuring a comprehensive and stable representation of its behavior. 

For this discussion, we denote the representation of a window $W$ of flows in the general form of $\mathbf{\mathrm{r}}_G(W; g)$. We also denote the of flows in a window $W$ by $n(W)$. Starting from a fixed anchor time, the algorithm computes the initial representation $\mathbb{\mathrm{r}}_G(W_0 ;g)$ for window $W_0$ of duration $L(W_0)=1$ day. This is used to initialize the reference representation $\mathbb{\mathrm{r}}_{ref}=\mathbb{\mathrm{r}}_G(W_0 ;g)$ and the reference number of flows $n_{ref}=W_0$. 

Afterwards, the algorithm proceeds in iterations, exponentially increasing the size of the observation window while keeping its start time fixed at the anchor time. At each iteration $i$, the relative increase in flow count compared to the reference is calculated as $\frac{n(W_i)-n_{ref}}{n_{ref}}$. If this increase exceeds a threshold (set at 0.5), the representation $\mathbb{\mathrm{r}}_G(W_i ;g)$ and its cosine similarity to $\mathbb{\mathrm{r}}_{ref}$ are computed. If the similarity exceeds a threshold $\theta$, then $\mathbb{\mathrm{r}}_G(W_i ;g)$ is exported as the service-level fingerprint. 

If the similarity threshold is not met, the reference representation and flow counts are updated with $\mathbb{\mathrm{r}}_G(W_i ;g)$ and $n(W_i)$, respectively, and the process continues. In cases where the relative increase in flow count does not exceed the threshold (\eg due to device inactivity, network outages, or measurement issues), the algorithm skips the current window, assuming insufficient new data. 

The maximum window size is capped at $2^6=64$ days (approximately two months). Note that the algorithm may terminate without exporting any fingerprint if the similarity condition is never satisfied. Finally, if convergence is achieved at some iteration $i_*$, the preceding window duration $L(W_{i_*-1})$ can be interpreted as an approximate seasonal period during which the device's network service usage stabilizes. This is inferred from the fact that extending the window to $W_{i_{*}}$, with a duration twice as long as that of $W_{i_{*}-1}$, does not significantly change the representation. A formal description of this procedure is provided in Algorithm 1 in Appendix \ref{App:exporter}, with further discussion of the implementation  design choices available in Appendix \ref{app:designChoices}.

\subsection{Convergence of Service-level Fingerprints}
\label{subsec:convergence}
\begin{figure*}[t!]
  \centering
  \begin{subfigure}{0.48\linewidth}
    \centering
    \includegraphics[width=0.88\linewidth]{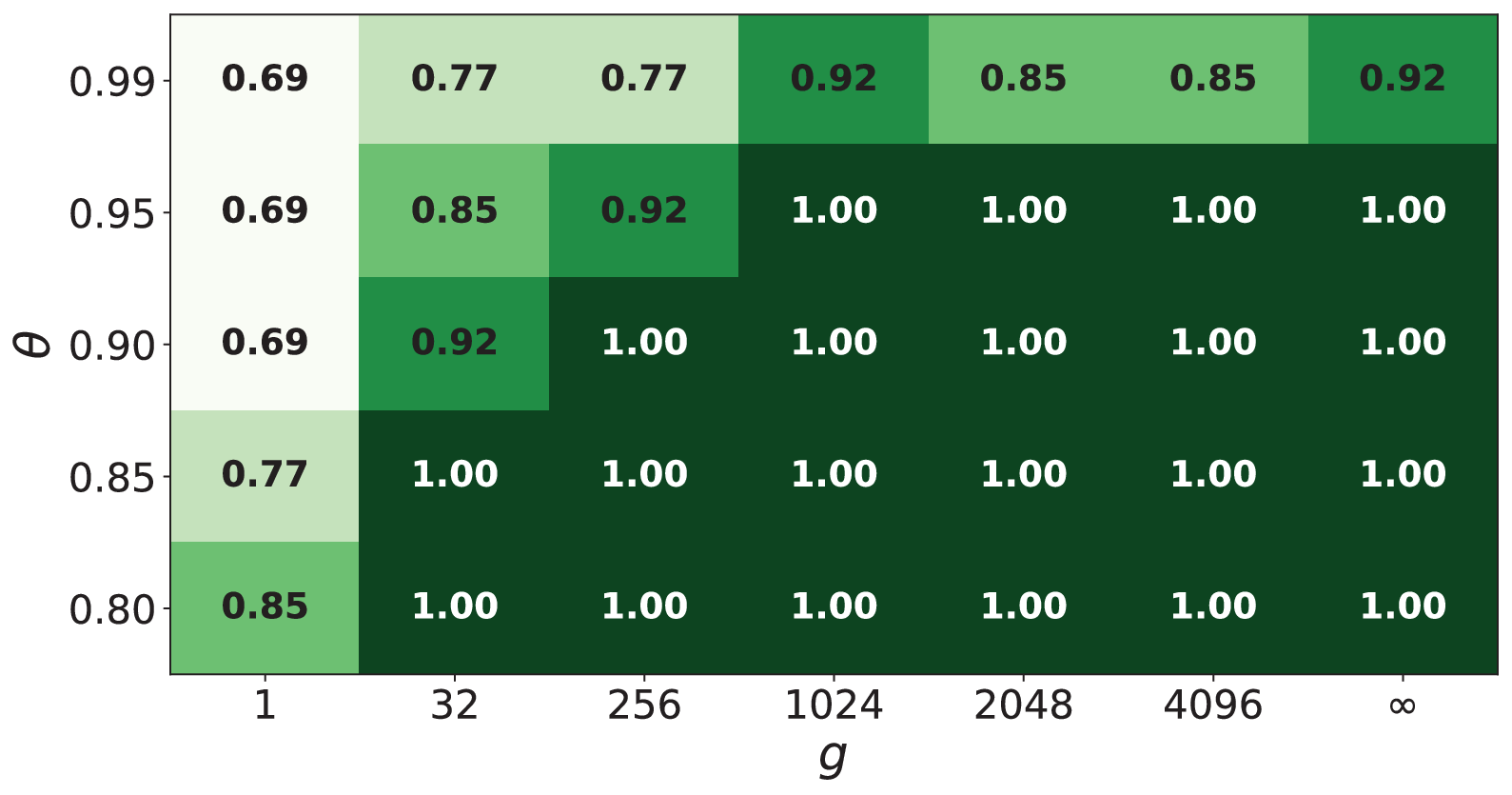}
    \caption{Fraction of fingerprinted devices.}
    \label{fig:convergence}
  \end{subfigure}
  \hfill
  \begin{subfigure}{0.48\linewidth}
    \centering
    \includegraphics[width=0.88\linewidth]{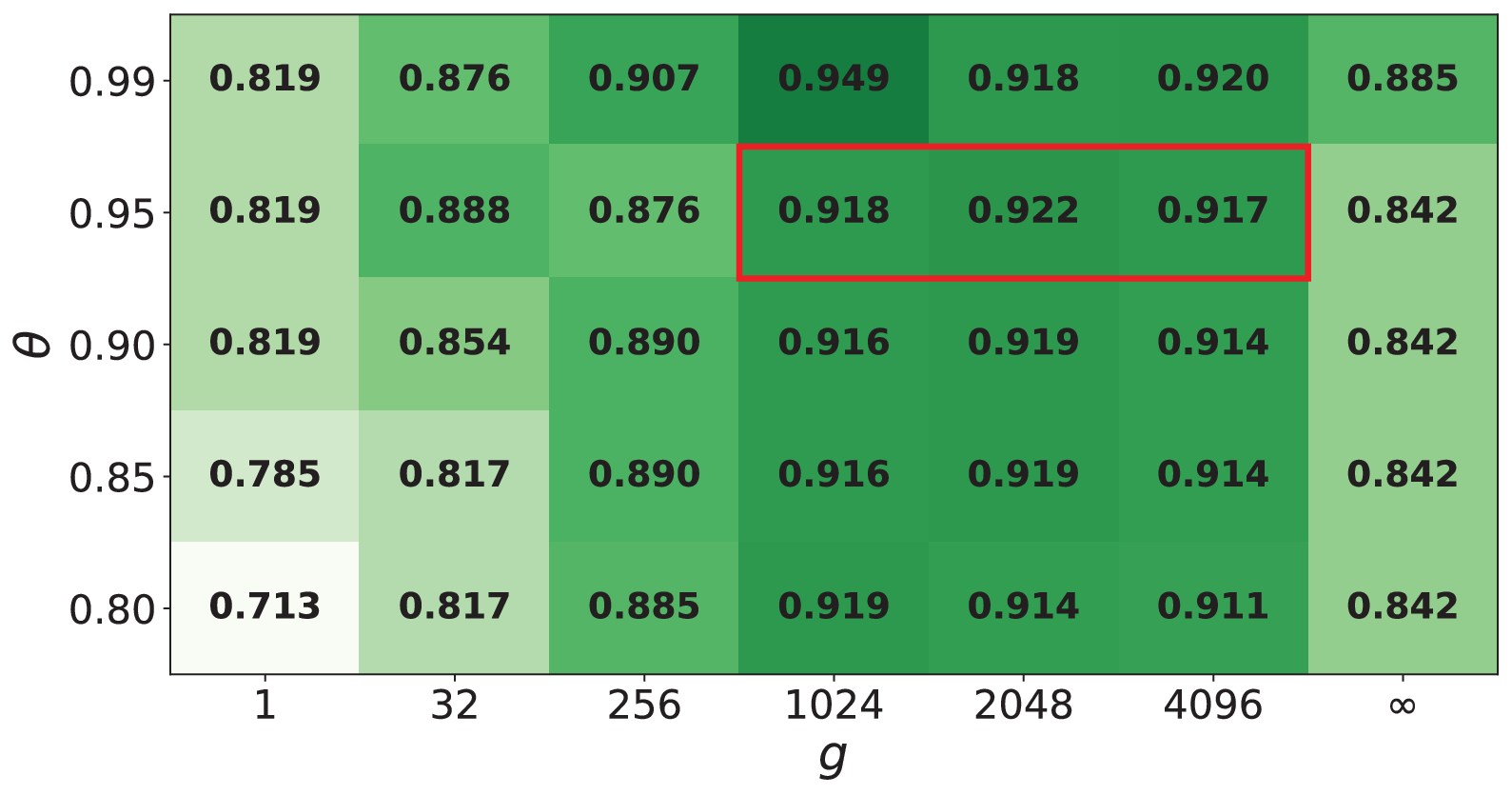}
    \caption{Average recurrence score of fingerprinted devices.}
    \label{fig:recurrence}
  \end{subfigure}
  \vspace{-3mm}
  \caption{Effect of threshold $\theta$ and granularity level $g$ on: (a) fingerprint convergence (fraction of fingerprinted devices), and (b) recurrence score. Higher $\theta$ and lower $g$ reduce convergence, while recurrence improves with stricter $\theta$ but degrades at extreme $g$ values.}  \label{fig:convergence_recurrence}
  \vspace{-4mm}
\end{figure*}

The operation of the fingerprint exporter, as explained above, depends on two key parameters: the similarity threshold $\theta$, and the granularity level $g$. A very stringent choice of $\theta$ may prevent the exporter from converging within the designated training data budget of 64 days. Conversely, as discussed in \S\ref{sec:contribution1}, non-stationarity of service usage can destabilize representations for certain devices, especially in the case of choosing a smaller $g$ combined with the presence of sporadic services. This may hinder the convergence of fingerprint export. To understand the effect of these parameters, we applied the fingerprint export procedure to traffic traces from 13 IoT devices in our testbed, starting from June 1st 2019. Fig. \ref{fig:convergence} shows the fraction of devices for which the exporter successfully converged under various settings. As expected, the convergence fraction decreases with increasing $\theta$, reflecting the strictness required for achieving the target cosine similarity between consecutive representations. Also, it can be seen that lower granularity levels (\ie smaller $g$ values) are typically associated with higher rates of non-convergence for devices, an effect consistent with earlier observations (\eg the QWatch Camera in \S\ref{sec:contribution1}).
Let us defer the discussion of insights from Fig.~\ref{fig:recurrence} to the next subsection.

\textbf{Durations of Service Usage Seasonalities:} 
As discussed in \S\ref{subsec:exporter}, the fingerprint exporter identifies a stable window of activity (interpreted as the device's seasonality period) for its service-level fingerprint. At an aggregate level, considering all exported fingerprints, over 90\% of the inferred periods are less than or equal to 8 days, suggesting a weekly pattern in network service usage. As expected, the distribution of inferred seasonality varies for different $(g,\theta)$ configurations. Across all combinations of $(g,\theta)$, the 80th percentile of inferred period remains at or below 8 days, except for $(g,\theta)=(1024, 0.99)$, where it rises to 16 days. These observations suggest that an 8-day window can be a reasonable upper bound for capturing IoT service usage patterns.

\subsection{Recurrence of Service-level Fingerprints}
\label{subsec:recurrenceScores}
So far, we have examined how the choice of granularity level in representing service usage compositions, together with the value of threshold $\theta$, affects the convergence of fingerprints. While a representation that enables convergence for a larger number of IoT devices is certainly desirable, convergence alone is not a sufficient criterion for success. Another critical factor is the recurrence quality of the exported fingerprint; how consistently it reappears in future observations and how robust it is to transient fluctuations in service usage. A high-quality fingerprint is essential to ensure the reliability and effectiveness of device identification.

To assess the \textit{recurrence} of a service-level fingerprint, we compute its cosine similarity with 18 consecutive 8-day windows following its export (after the latest fingerprint's convergence at 64 days, there are 18 complete 8-day windows in 2019).
Based on the closing discussion in \S\ref{subsec:convergence}, the duration of each window is set at 8 days, as this is the typical period during which a device exhibits its full range of network service usage. 
In Fig. \ref{fig:recurrence}, we show the average of mean recurrence scores of the exported fingerprints across devices. As expected, greater strictness (\ie higher values of $\theta$) mostly results in higher quality fingerprints with better recurrence. However, this comes at the cost of reduced coverage, since stricter thresholds may prevent the exporter from capturing fingerprints for some devices (as seen in the top row of Fig. \ref{fig:convergence}). Additionally, both overly coarse-grained ($g\to1$) and overly fine-grained ($g=\infty$) representations result in poor recurrence. Instead, we observe that an  \textit{intermediate granularity level} often achieves the best balance, as seen by the trends in Fig.~\ref{fig:recurrence}. 

The two matrices in Figs.~\ref{fig:convergence} and ~\ref{fig:recurrence} can also guide the selection of optimal parameters $\theta$ and $g$ for fingerprinting devices and later using those fingerprints for detection. For example, from Fig.~\ref{fig:convergence} it can be seen that $\theta=0.95$ is the highest convergence threshold for which we have full coverage at least for some values of $g$ (specifically, $\{1024, 2048, 4096, \infty\}$). Note that higher values of $\theta$ are generally preferable,  since stricter convergence criteria during fingerprint export reduce the likelihood of fingerprint overlap, thus improving device identification. Now, according to Fig.~\ref{fig:recurrence}, at $\theta=0.95$, intermediate $g$ values of $\{1024, 2048, 4096\}$ yield the best recurrence quality. Therefore, the highlighted rectangle in Fig.~\ref{fig:recurrence}, seems to be a suitable operating point.

\section{Device Identification}
\label{sec:contribution3}
So far, we have explored how IoT devices differ in their service usage patterns, developed a method to represent these patterns as service-level fingerprints, and analyzed how different representation choices and threshold settings impact fingerprint quality. In this section, we demonstrate how the exported fingerprints can be used to accurately identify IoT devices based on their network traffic.

We start our evaluations in a \textbf{closed-set} scenario, where the test data corresponds to the same 13 devices that were fingerprinted. The inference task involves classifying the traffic generated by the devices active within each inference window. 
Each inference window spans 8 days, a duration during which devices typically exhibit their full service usage behaviors (as discussed in \S\ref{sec:contribution2}).
These windows are generated using a sliding window approach, starting on January 1st, 2020, with a one-day shift at each step. This results in a total of 359 inference windows per device. Following the closed-set evaluation, we extend our analysis to an \textbf{open-set} classification scenario by incorporating test data from 22 previously unseen devices. In this section, we only focus on configurations where $\theta=0.95$ and $g\in\{1024, 2048, 4096, \infty\}$ under which all devices were successfully fingerprinted with a reasonably strict convergence criterion (see Fig. \ref{fig:convergence}).


\subsection{Closed-Set Classification} 
Given an inference window $W_*$ of flow records for each device, and a set
of exported fingerprint representations, we classify the device by identifying the fingerprint that has the maximum cosine similarity to the representation derived from $W_*$. However, this method requires a mechanism for conflict resolution. Specifically, two of our 13 IoT classes, Nature Remo and Qrio Hub, exclusively use a single service (TCP/443) throughout their operation. As a result, the two devices are indistinguishable in the test phase due to the cosine similarity of 1 to both their own and each other's exported fingerprints. To address this, we incorporate the information conveyed by the size of the representation vectors. 
In the event of a prediction conflict, each candidate device's fingerprint is linearly scaled and capped at size  $g$, based on the ratio between the inference window and the original fingerprint window. We then compute the ratio of the fingerprint's norm-1 to that of the inference window's representation. 
The candidate device is selected by minimizing the squared logarithmic deviation of this ratio from 1.
This approach penalizes significant deviations from expected traffic volume and helps differentiate otherwise identical fingerprints. Macro-averaged precision and recall metrics for this classifier are presented in the first row of Table \ref{tab:metrics}. It can be seen that $g=2048$ provides the best performance, achieving a precision of $0.74$ and a recall of $0.83$, while the poorest performance is observed at $g=\infty$. This is aligned with our earlier observations regarding the recurrence quality of fingerprints across various values of $g$ (see Fig. \ref{fig:recurrence}). 


\begin{table}[t!]
  \caption{Macro-averaged ``Precision''/``Recall'' across different classification settings.}
  \vspace{-3mm}
  \label{tab:metrics}
  \centering
  \begin{adjustbox}{max width=\linewidth}
  \begin{tabular}{lccccccc}
    \toprule
    \textbf{Classification setting} & 
    \textbf{$g=1024$} & \textbf{$g=2048$} & \textbf{$g=4096$} & \textbf{$g=\infty$} \\
    \midrule
    \textbf{Closed-set (original) } & 
    $0.73/0.81$ & \textbf{0.74/0.83} & $0.68/0.75$ & $0.49/0.54$ \\
    \textbf{Closed-set  (augmented)} & 
    $0.87/0.88$ & \textbf{0.98/0.97} & $0.94/0.96$ & $0.91/0.90$ \\
    \textbf{Open-set (augmented)} & 
    $0.77/0.80$ & \textbf{0.79/0.86} & $0.76/0.81$ & $0.75/0.82$ \\
    \bottomrule
  \end{tabular}
  \end{adjustbox}
  \vspace{-5mm}
\end{table}

Examining the case of $g=2048$, we identified three devices as the main contributors to the imperfect performance of our classifier: JVCKENWOOD Hub, Line Clova Wave and Planex Camera Outdoor. The first two were never classified correctly, while the Planex Camera achieved a recall of 86\%. In contrast, the remaining  classes,  were  classified with recall values of 98\% or higher.
By analyzing changes of recurrence scores for these devices, we found out that certain recurrence windows experienced fairly low similarity scores. This suggests the emergence of behavioral patterns  not fully captured in the initial fingerprint representations. In other words, although IoT devices may exhibit converging and stable service-level behaviors over shorter time periods (\eg a couple of weeks), additional aspects of their service usage may emerge only over longer time durations (\eg a couple of months). 
To address this limitation, we introduce an extension to our fingerprinting method. For each device, once an initial fingerprint is exported at a specific configuration $(g, \theta)$, we continuously monitor the cosine similarity between subsequent data windows and all existing fingerprints for that device. Initially, the comparison is made only with the original fingerprint. If the similarity score drops below $\theta$, the fingerprint exporter (as defined in \S\ref{sec:contribution2}) is reactivated. If convergence occurs, the resulting fingerprints is added the device's fingerprint pool as a new variant capturing an alternative behavioral pattern. If the convergence is not achieved,  monitoring resumes with the next data window. This process continues until the training data (the end of 2019) is fully exhausted. 

After this augmentation, we observe that $g=\infty$ leads to the highest average number of additional fingerprints per device at 2.8, while $g=2048$ results in the fewest, at 2.1. Although the goal of augmenting fingerprints is to improve classification performance, a higher number of fingerprints may increase the chance of overlap and misclassification. In this regard, the lower count associated with $g=2048$ is more desirable. Additionally, representation granularity itself plays an important role in class separation. The second row of Table \ref{tab:metrics} shows performance at different granularity values, where $g=2048$ achieves best performance with precision of $0.98$ and recall of $0.97$. The corresponding confusion matrix for this setting is presented in Fig. \ref{fig:CM_multi_veil_closed_set} in Appendix \ref{app:sec:CMs}. 

\subsection{Open-Set Classification}
We now consider the case of open-set classification, where traffic of 22 previously unseen device types is introduced to the system. In such scenarios, the classifier is expected to abstain from making incorrect predictions by labeling unfamiliar traffic as ``UNKNOWN''. 
This can be achieved by incorporating a measure of confidence and setting a threshold to discard predictions with low confidence. In practice, the maximum cosine similarity between an unseen device type and known fingerprints tends to be low. To leverage this, we estimate a confidence threshold based on the typical similarity values observed during training period. 
Specifically, we examined 26 consecutive and non-overlapping 8-day windows in training period and computed the maximum cosine similarity of the device's representation to all its exported fingerprints in each window. From these, we calculated the mean ($\mu$) and standard deviation ($\sigma$) of the similarity scores. 

During inference, predictions with a maximum cosine similarity below $\mu -3\sigma$ are labeled as ``UNKNOWN'' and subsequently discarded. The performance of the classifier under this open-set setting is summarized in the third row of Table \ref{tab:metrics}, evaluated at different values of $g$. Once again, best performance is observed at $g=2048$, confirming its robustness across different settings. The corresponding confusion matrix is presented in Fig.~\ref{fig:CM_multi_veil_open_set} in Appendix \ref{app:sec:CMs}. 
For consistency, we aggregate all unseen classes into a single class ``UNKNOWN'',  which is treated like any other class in the macro-averaged precision and recall metrics.
Our classifier successfully discards 10 out of the 22 unseen classes more than 99\% of the time and discards three more over 50\% of the time. However, it struggles to consistently identify the remaining nine as unseen.
About 80\% of these accepted misclassifications are concentrated in two known classes: JVCKENWOOD IP Camera (49\%) and Line Clova Wave (31\%). 
When UNKNOWN predictions are treated as filtered, recall remains stable compared to the closed-set scenario. However, precision declines for several classes: dropping by 0.2 to 0.5 for four classes, and by as much as 0.90 for the JVCKENWOOD IP Camera.
\section{Related Work}
\textbf{ML-based IoT traffic classification:} Extensive research has explored ML models using packet-level \cite{Sivanathan:TMC19, Miettinen:ICDCS17, DEFT} or flow-level \cite{Meidan:CS20, Pashamokhtari:LCN21}  features for IoT traffic classification. 
However, these methods face key limitations. They often depend on fine-grained timer or counter-based features that are computationally expensive at scale and prone to telemetry errors \cite{Rexford:CSUR25, Sekar:SOSR21, Trammell:PAM11, Paxson:IMC04}. Performance degradation over time is a well-documented issue \cite{Maali:NDSS25, Arman:IoTJ23, Feamster:CONEXT23, Shahbaz:OSDI24, Malekghaini:CompNet2023, Kolcun:TMA2021, Azizi:Sigcomm24, Anderson:SIGKDD2017}, with some IoT classifiers showing decays within weeks  \cite{Kolcun:TMA2021}. Additional concerns include limited interpretability \cite{Maali:NDSS25, Nascita:CST24}, challenges in handling unknown classes in open-set conditions \cite{Paxson:SP10, Wang:TIFS25, Jorgensen:TAI24, Yang:TNSM21, Zhang:INFOCOM20}, and ineffective use of traffic types (\eg NTP) for device identification \cite{Pashamokhtari:LCN21}. Overly relying on ML-based methods may also divert attention from traffic modeling approaches that could complement ML methods for better accuracy and validation.

\textbf{IoT classification via explicit traffic modeling:} Beyond packet-level or flow metadata methods, directly modeling IoT network behavior has also been used for device identification, aligning with our approach. Closely related works include \cite{20TDSCmud, IMC2020Haystack, 18iciafsIoT}.
In \cite{20TDSCmud}, MUD profiles (\ie lists of transport-layer protocols, port numbers, and endpoints) are structured as trees for fingerprinting and compared dynamically to observed traffic.
The work in \cite{IMC2020Haystack} employs a bag-of-domains approach, matching devices based on the domains they have contacted, but struggles when DNS payloads are unavailable, as mapping IP addresses to domains is complex. The authors of \cite{18iciafsIoT} focused on transport-layer services exposed by IoT devices via active port scanning, unlike passive observation of services they have consumed. Other explicit modeling approaches include \cite{Marchal:JSAC19}, which models non-user flows as periodic time series, and \cite{IoTFinder}, which uses NLP on DNS queries but relies on payloads and only covers  DNS traffic. 
Additional studies, such as \cite{Choffnes:IMC23} and \cite{Hamza:SOSR19}, model IoT behaviors through periodic events and user-triggered events or cluster transitions via state machines. 
Finally, work in \cite{Feamster:NDSS16} links changes in IoT traffic rate to user interactions.


\section{Conclusion}
In this work, we introduced a lightweight and interpretable approach to IoT device fingerprinting which focuses on the set of network services that devices routinely access, rather than relying on complex, low-level packet or flow features. We showed that service usage patterns are relatively stable and distinctive across different types of IoT devices, and developed a generalized method to represent these patterns with configurable granularity. Our fingerprint extraction procedure enables automated generation of device-specific service-level fingerprints and successfully identifies the time window over which device behavior converges. We evaluated the efficacy of our fingerprints for 13 representative IoT device classes by applying them to one year of flow data (testing period) in closed-set and open-set scenarios. Our results demonstrate that known (seen) devices can be classified with high accuracy, while previously unseen devices can be identified with reasonably reliable predictions.

\balance
\bibliographystyle{ACM-Reference-Format}
\bibliography{ServiceFingerprintIoT}

@ARTICLE{DEFT,
  author={Thangavelu, Vijayanand and Divakaran, Dinil Mon and Sairam, Rishi and Bhunia, Suman Sankar and Gurusamy, Mohan},
  journal={IEEE IoTJ}, 
  title={DEFT: A Distributed IoT Fingerprinting Technique}, 
  year={2019},
  volume={6},
  number={1},
  pages={940-952}
}

@INPROCEEDINGS{Feamster:NDSS16,
  author = {Apthorpe, Noah and Reisman, Dillon and Feamster, Nick},
  booktitle={Proc. NDSS}, 
  title={{Poster: A Smart Home is No Castle: Privacy Vulnerabilities of Encrypted IoT Traffic}}, 
  year={2016},
  month = {Feb},
  address = {San Diego, CA, USA}
}

@misc{RFC8520,
    series =    {Request for Comments},
    number =    8520,
    howpublished =  {RFC 8520},
    publisher = {RFC Editor},
    doi =       {10.17487/RFC8520},
    url =       {https://www.rfc-editor.org/info/rfc8520},
    author =    {Eliot Lear and Ralph Droms and Dan Romascanu},
    title =     {{Manufacturer Usage Description Specification}},
    pagetotal = 60,
    year =      2019,
    month =     {Mar},
}

@INPROCEEDINGS{IoTFinder,
  author={Perdisci, Roberto and others},
  booktitle={Proc. IEEE EuroS\&P}, 
  title={{IoTFinder: Efficient Large-Scale Identification of IoT Devices via Passive DNS Traffic Analysis}}, 
  year={2020},
  month = {Sep},
  address = {Virtual Event}
}

@inproceedings{Hamza:SOSR19,
author = {Hamza, Ayyoob and Gharakheili, Hassan Habibi and Benson, Theophilus A. and Sivaraman, Vijay},
title = {Detecting Volumetric Attacks on loT Devices via SDN-Based Monitoring of MUD Activity},
year = {2019},
month = {Apr},
booktitle = {Proc. ACM SOSR},
location = {San Jose, CA, USA}
}

@inproceedings{Choffnes:IMC23,
author = {Hu, Tianrui and Dubois, Daniel J. and Choffnes, David},
title = {BehavIoT: Measuring Smart Home IoT Behavior Using Network-Inferred Behavior Models},
year = {2023},
month = {Oct},
booktitle = {Proc. ACM IMC},
address = {Montreal QC, Canada}
}

@ARTICLE{Marchal:JSAC19,
  author={Marchal, Samuel and Miettinen, Markus and Nguyen, Thien Duc and Sadeghi, Ahmad-Reza and Asokan, N.},
  journal={IEEE JSAC}, 
  title={AuDI: Toward Autonomous IoT Device-Type Identification Using Periodic Communication}, 
  year={2019},
  volume={37},
  number={6},
  pages={1402-1412},
}

@inproceedings{Shahbaz:OSDI24,
author =  {Zhang, Qizheng and Imran, Ali and Bardhi, Enkeleda and Swamy, Tushar and Zhang, Nathan and Shahbaz, Muhammad and Olukotun, Kunle},
title = {CARAVAN: Practical Online Learning of In-network ML Models with Labeling Agents},
year = {2024},
month = {July},
booktitle = {Proc. USENIX OSDI},
address = {Santa Clara, CA, USA},
}

@inproceedings{Kolcun:TMA2021,
  author  = {Roman Kolcun and
  Diana Andreea Popescu and
  Vadim Safronov and
  Poonam Yadav and
  Anna Maria Mandalari and
  Richard Mortier and
  Hamed Haddadi},
  title        = {{Revisiting IoT Device Identification}},
  booktitle    = {Proc. IFIP TMA},
  address = {Virtual Event},
  year  = {2021},
}

@article{Malekghaini:CompNet2023,
  author       = {Navid Malekghaini and
  Elham Akbari and
  Mohammad A. Salahuddin and
  Noura Limam and
  Raouf Boutaba and
  Bertrand Mathieu and
  Stephanie Moteau and
  St{\'{e}}phane Tuffin},
title        = {{Deep Learning for Encrypted Traffic Classification in the Face of
  Data Drift: An Empirical Study}},
  journal   = {Elsevier Computer Networks},
  year   = {2023},
  volume={225},
  pages={109648-},
}

@inproceedings{Anderson:SIGKDD2017,
author = {Anderson, Blake and McGrew, David},
title = {Machine Learning for Encrypted Malware Traffic Classification: Accounting for Noisy Labels and Non-Stationarity},
year = {2017},
booktitle = {Proc. ACM SIGKDD},
address = {Halifax, NS, Canada}
}

@inproceedings{Azizi:Sigcomm24,
author = {Azizi, Shayan and Okui, Norihiro and Nakahara, Masataka and Kubota, Ayumu and Batista, Gustavo and Gharakheili, Hassan Habibi},
title = {Poster: Understanding and Managing Changes in IoT Device Behaviors for Reliable Network Traffic Inference},
year = {2024},
booktitle = {Proc. SIGCOMM Posters and Demos},
address = {Sydney, NSW, Australia},
}

@article{Feamster:CONEXT23, 
author = {Liu, Shinan and Bronzino, Francesco and Schmitt, Paul and Bhagoji, Arjun Nitin and Feamster, Nick and Crespo, Hector Garcia and Coyle, Timothy and Ward, Brian}, 
title = {{LEAF: Navigating Concept Drift in Cellular Networks}}, 
year = {2023},  
publisher = {ACM}, 
volume = {1}, 
number = {CoNEXT2}, 
journal = {PACMNET}, 
month = {Sep}, 
numpages = {24}}

@inproceedings{Maali:NDSS25,
author = {Eman Maali and Omar Alrawi and Julie McCann},
title = {Evaluating Machine Learning-Based IoT Device Identification Models for Security Applications},
year = {2025},
month = {Feb},
booktitle = {Proc. NDSS},
address = {San Diego, CA, USA}
}

@ARTICLE{Nascita:CST24,
  author={Nascita, Alfredo and Aceto, Giuseppe and Ciuonzo, Domenico and Montieri, Antonio and Persico, Valerio and Pescapé, Antonio},
  journal={IEEE Communications Surveys \& Tutorials}, 
  title={A Survey on Explainable Artificial Intelligence for Internet Traffic Classification and Prediction, and Intrusion Detection}, 
  year={2024}
 }

@ARTICLE{Yang:TNSM21,
  author={Yang, Lixuan and Finamore, Alessandro and Jun, Feng and Rossi, Dario},
  journal={IEEE TNSM}, 
  title={Deep Learning and Zero-Day Traffic Classification: Lessons Learned From a Commercial-Grade Dataset}, 
  year={2021},
  volume={18},
  number={4},
  pages={4103-4118},
  }

@INPROCEEDINGS{Zhang:INFOCOM20,
  author={Zhang, Jielun and others},
  booktitle={Proc. IEEE INFOCOM}, 
  title={Autonomous Unknown-Application Filtering and Labeling for DL-based Traffic Classifier Update}, 
  year={2020},
  month={Jul},
  address={Toronto, ON, Canada}
  }

@ARTICLE{Jorgensen:TAI24,
  author={Jorgensen, Steven and Holodnak, John and Dempsey, Jensen and de Souza, Karla and Raghunath, Ananditha and Rivet, Vernon and DeMoes, Noah and Alejos, Andrés and Wollaber, Allan},
  journal={IEEE TAI}, 
  title={Extensible Machine Learning for Encrypted Network Traffic Application Labeling via Uncertainty Quantification}, 
  year={2024},
  volume={5},
  number={1},
  pages={420-433},
}

@ARTICLE{Wang:TIFS25,
  author={Wang, Xueman and Wang, Yipeng and Lai, Yingxu and Hao, Zhiyu and Liu, Alex X.},
  journal={IEEE TIFS}, 
  title={Reliable Open-Set Network Traffic Classification}, 
  year={2025},
  volume={20},
  pages={2313-2328}}

@INPROCEEDINGS{Paxson:SP10,
  author={Sommer, Robin and Paxson, Vern},
  booktitle={Proc. IEEE S\&P}, 
  title={Outside the Closed World: On Using Machine Learning for Network Intrusion Detection}, 
  year={2010},
  month={May},
  address={Oakland, CA, USA}
}

@inproceedings{Paxson:IMC04,
author = {Paxson, Vern},
title = {Strategies for Sound Internet Measurement},
year = {2004},
month = {Oct},
booktitle = {Proc. ACM IMC},
address = {Taormina, Sicily, Italy}
}

@inproceedings{Trammell:PAM11,
author = {Trammell, Brian and Tellenbach, Bernhard and Schatzmann, Dominik and Burkhart, Martin}
,
title = {Peeling Away Timing Error in NetFlow Data},
year = {2011},
booktitle = {Proc. PAM},
address = {Atlanta, GA, USA}
}

@article{Rexford:CSUR25,
author = {Landau-Feibish, Shir and Liu, Zaoxing and Rexford, Jennifer},
title = {Compact Data Structures for Network Telemetry},
year = {2025},
issue_date = {August 2025},
publisher = {Association for Computing Machinery},
volume = {57},
number = {8},
journal = {ACM CSUR},
month = {Mar},
articleno = {191},
numpages = {31}
}

@inproceedings{Sekar:SOSR21,
author = {Namkung, Hun and Kim, Daehyeok and Liu, Zaoxing and SekaR, Vyas and Steenkiste, Peter},
title = {Telemetry Retrieval Inaccuracy in Programmable Switches: Analysis and Recommendations},
year = {2021},
month = {Oct},
booktitle = {Proc. SOSR},
address = {Virtual Event, USA}
}

@misc{Feamster:Blog17,
  author       = {Nick Feamster},
  title        = {Mitigating the Increasing Risks of an Insecure Internet of Things},
  year         = {2017},
  month        = {Feb},
  url          = {https://blog.citp.princeton.edu/2017/02/18/mitigating-the-increasing-risks-of-an-insecure-internet-of-things/},
  note         = {Accessed: 2025-06-04},
  publisher    = {Princeton University Center for Information Technology Policy},
}

@INPROCEEDINGS{Miettinen:ICDCS17,
  author={Miettinen, Markus and Marchal, Samuel and Hafeez, Ibbad and Asokan, N. and Sadeghi, Ahmad-Reza and Tarkoma, Sasu},
  booktitle={Proc. IEEE International Conference on Distributed Computing Systems (ICDCS)}, 
  title={{IoT SENTINEL: Automated Device-Type Identification for Security Enforcement in IoT}}, 
  year={2017},
  address={Atlanta, GA, USA}
}

@article{Arman:IoTJ23,
  author = {Arman Pashamokhtari and
  Norihiro Okui and
  Masataka Nakahara and
  Ayumu Kubota and
  Gustavo Batista and
  Hassan Habibi Gharakheili},
  title        = {{Dynamic Inference From IoT Traffic Flows Under Concept Drifts in Residential
  ISP Networks}},
  journal      = {{IEEE} IoTJ},
  volume       = {10},
  number       = {17},
  pages        = {15761--15773},
  year         = {2023}
}

@INPROCEEDINGS{Pashamokhtari:LCN21,
  author={Pashamokhtari, Arman and Okui, Norihiro and Miyake, Yutaka and Nakahara, Masataka and Gharakheili, Hassan Habibi},
  booktitle={Proc. IEEE LCN}, 
  title={{Inferring Connected IoT Devices from IPFIX Records in Residential ISP Networks}}, 
  year={2021},
  address={Virtual Event}
}

@article{Meidan:CS20,
  title={{A Novel Approach for Detecting Vulnerable IoT Devices Connected
Behind a Home NAT}},
  author = {Meidan, Yair and Sachidananda, Vinay and Peng, Hongyi and Sagron, Racheli and Elovici, Yuval and Shabtai, Asaf},
  journal={Computers \& Security},
  year={2020},
  volume={91},
  pages={101968}
}

@article{Sivanathan:TMC19,
  title={{Classifying IoT Devices in Smart Environments Using Network Traffic Characteristics}},
  author={Arunan Sivanathan and Hassan Habibi Gharakheili and Franco Loi and Adam Radford and Chamith Wijenayake and Arun Vishwanath and Vijay Sivaraman},
  journal={IEEE TMC},
  year={2019},
  volume={18},
  pages={1745-1759}
}

@misc{techtarget:CSAM,
  author       = {{Andrew Froehlich}},
  title        = {CyberSecurity Asset Management (CSAM)},
  year         = 2023,
  url          = {https://www.techtarget.com/searchsecurity/definition/cybersecurity-asset-management-CSAM},
  note         = {Accessed: 2025-06-04},
  publisher    = {TechTarget},
}

@techreport{Gartner:IoTSec,
  author       = {Gartner Research},
  title        = {IoT Security Primer: Challenges and Emerging Practices},
  year         = {2025},
  url          = {https://www.gartner.com/en/doc/iot-security-primer-challenges-and-emerging-practices}
}

@inproceedings{IMC2020Haystack,
author = {Saidi, Said Jawad and Mandalari, Anna Maria and Kolcun, Roman and Haddadi, Hamed and Dubois, Daniel J. and Choffnes, David and Smaragdakis, Georgios and Feldmann, Anja},
title = {{A Haystack Full of Needles: Scalable Detection of IoT Devices in the Wild}},
year = {2020},
month = {Oct},
booktitle = {Proc. IMC},
address = {Virtual Event, USA},
}

@ARTICLE{20TDSCmud,
	author={Hamza, Ayyoob and others},
	journal={IEEE TDSC}, 
	title={{Verifying and Monitoring IoTs Network Behavior Using MUD Profiles}}, 
	year={2020},
	month={May},
	volume={19},
	number={1},
	pages={1-18},
}

@INPROCEEDINGS{18iciafsIoT,
	author={Sivanathan, Arunan and Habibi Gharakheili, Hassan and Sivaraman, Vijay},
	booktitle={Proc. IEEE ICIAfS}, 
	title = {{Can We Classify an IoT Device using TCP Port Scan?}},
	year={2018},
	month={Dec},
	address={Colombo, Sri Lanka},
}

@String{Computing = "Computing" }

@String{Computer = "{IEEE} Computer" }

\clearpage
\appendix

\clearpage
\appendix

\begin{figure*}[t!]
	\centering
	\begin{subfigure}{0.40\linewidth}
		\centering
		\includegraphics[width=\linewidth]{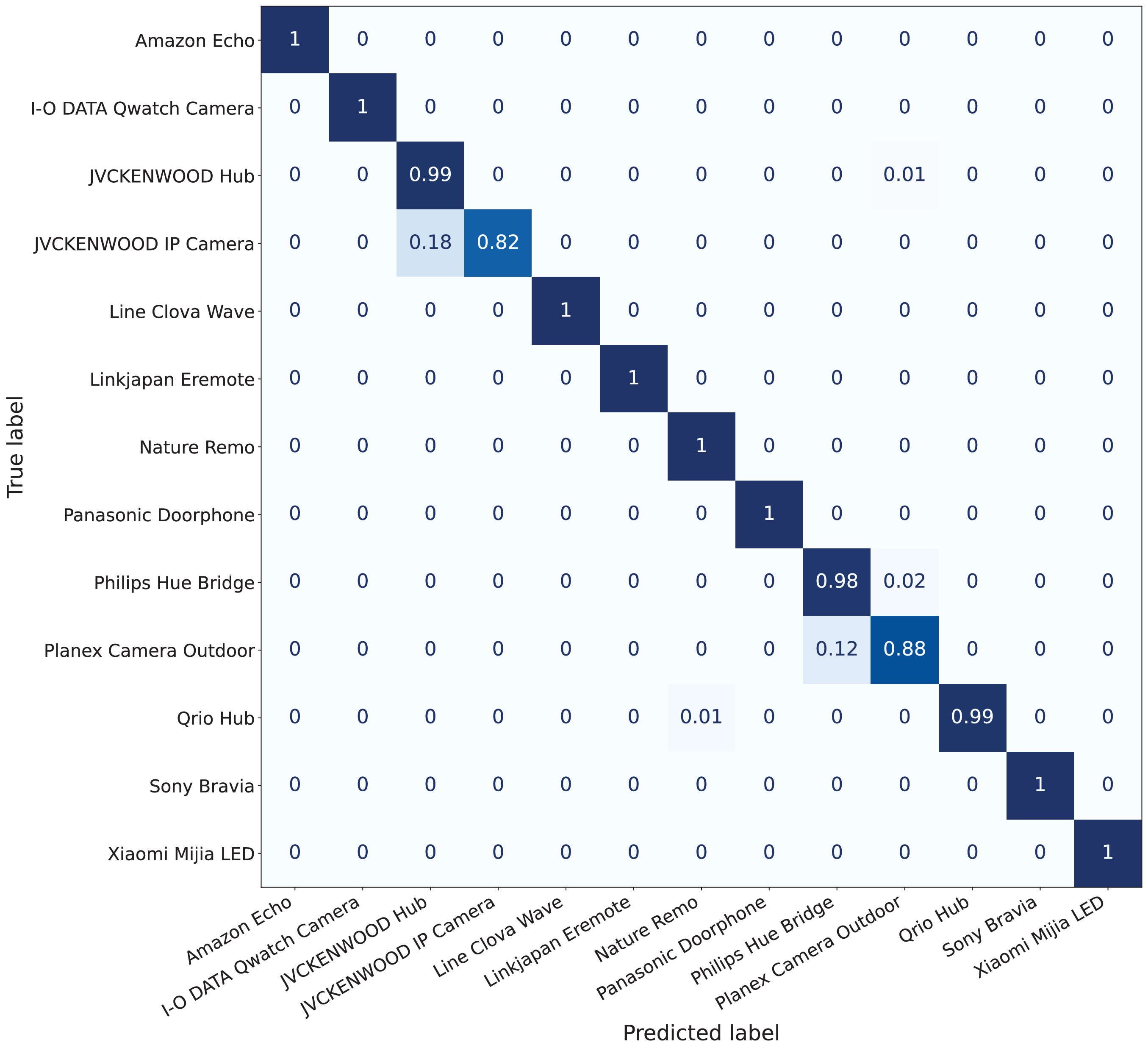}
		\caption{Closed-set classification.}\label{fig:CM_multi_veil_closed_set}
	\end{subfigure}
	\hfill
	\begin{subfigure}{0.40\linewidth}
		\centering
		\includegraphics[width=\linewidth]{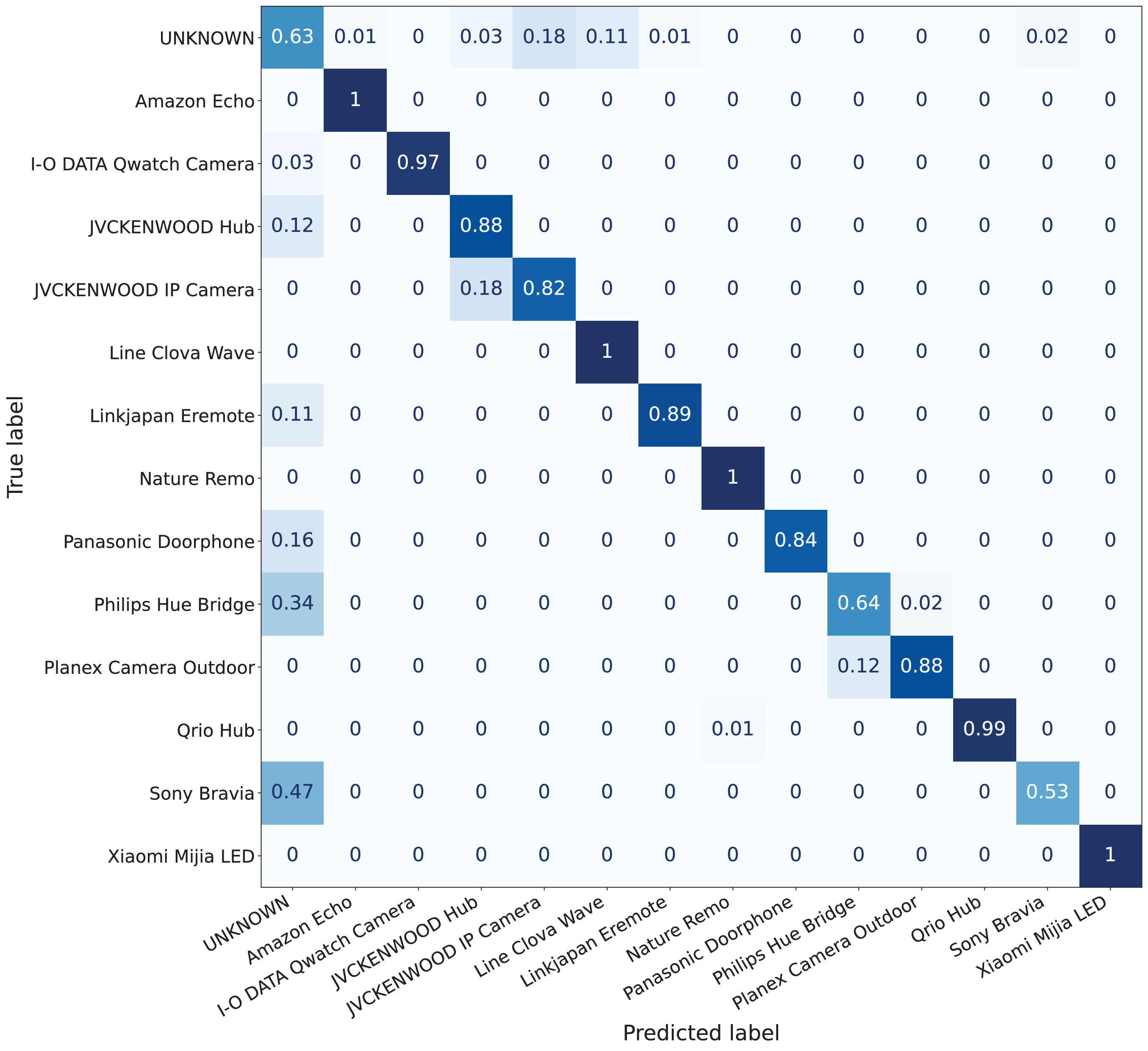}
		\caption{Open-set classification.}\label{fig:CM_multi_veil_open_set}
	\end{subfigure}
	\caption{Confusion matrices for classification using the augmented fingerprints pool at $(g, \theta)=(2048, 0.95)$ (a) closed-set scenario, (b) open-set scenario.}
	\label{fig:CM}
	\vspace{-4mm}
\end{figure*}

\section{Fingerprint Exporter}
\label{App:exporter}

The fingerprint export procedure (discussed in \S\ref{subsec:exporter}) is formally summarized in Algorithm 1, below.
\begin{algorithm}
\caption{Service-level Fingerprint Exporter}

\begin{algorithmic}[1]
\label{Alg:exporter}
\State \textbf{Input:} Anchor time $t_0$; initial window size $L_0$; maximum iterations $i_{max}$; similarity threshold $\theta$; minimum flow count growth rate $\delta$.
\State \textbf{Output:} Fingerprint $\mathbf{\mathrm{r}}_G(W_i; g)$ or \textit{``Did not converge''}
\State $W_0 \leftarrow$ flow records between $t_0$ and $ t_0+L_0$
\State $n_{ref} \leftarrow flow\_count(W_0)$
\State Compute $\mathbf{\mathrm{r}}_G(W_0; g)$
\State $\mathbf{\mathrm{r}}_{ref} \leftarrow \mathbf{\mathrm{r}}_G(W_0; g)$
\For{$i = 1$ to $i_{\max}$}
    \State $W_i \leftarrow$ flow records between $t_0$ and $ t_0+2^iL_0$
    \If{$flow\_count(W_i) - n_{ref} > \delta\cdot n_{ref}$} 
        \State Compute $\mathbf{\mathrm{r}}_G(W_i; g)$
        \State $n_{ref}\leftarrow flow\_count(W_i)$
        \If{$\lVert\mathrm{r}_{ref}\rVert_1 >0$}  \Comment{To handle the case of empty $W_0$}      
            \If{$s\bigl(\mathbf{\mathrm{r}}_G(W_i; g), \mathbf{\mathrm{r}}_{ref}\bigr) > \theta$} \Comment{Cosine similarity check}
                \State \Return $\mathbf{\mathrm{r}}_G(W_i; g)$
            \EndIf    
        \EndIf
        \State $\mathbf{\mathrm{r}}_{ref} \leftarrow \mathbf{\mathrm{r}}_G(W_i; g)$
    \EndIf
\EndFor
\State \Return \textit{``Did not converge''}
\end{algorithmic}
\end{algorithm}

\section{Fingerprint Exporter Design Choices}
\label{app:designChoices}
In the absence of prior knowledge about device behavior, using exponentially growing window sizes offers a relatively efficient search for identifying the time span over which a device reveals its full fingerprint. The choice of using an exponent of $2$ ensures that each new window introduces as much additional data as the previous, smaller window contained (except in cases where a window is skipped due to insufficient new flow data). 

As discussed in the paper, if the similarity between two successive representations is high, it is reasonable to assume that the smaller window may correspond to a natural ``period'' of the device's network usage behavior. If convergence occurs at iteration $i$, that is, $s\bigl(\mathbf{\mathrm{r}}(W_{i};g),\mathrm{r}_{ref}\bigr) \geq \theta$ (see Algorithm 1 above), then $\mathbf{\mathrm{r}}(W_{i};g)$, instead of $\mathrm{r}_{ref}$, is exported as the device fingerprint. This allows the final fingerprint to be derived from a larger and more representative data sample. 
The window size sequence $\{1, 2, 4, 8, 16, 32, 64\}$ is intuitive and efficient, capturing typical periodicities in device behavior, such as daily ($1\rightarrow2$), weekly ($8\rightarrow16$), and monthly ($32\rightarrow64$) patterns.

\section{Confusion Matrices}
\label{app:sec:CMs}
The confusion matrices of the best-performing classifiers in closed-set and open-set scenarios are shown in Fig. \ref{fig:CM_multi_veil_closed_set} and Fig. \ref{fig:CM_multi_veil_open_set}, respectively. The matrices are row-normalized. As illustrated in Fig. \ref{fig:CM_multi_veil_open_set}, in 63\% of cases, unseen devices are correctly classified as UNKNOWN, while 29\% of these instances are misclassified into just two known classes. On the other hand, the classifier has become more conservative by rejecting correct predictions for some seen classes, including Sony Bravia (47\% rejection), Philips Hue Bridge (34\%), Panasonic Doorphone (16\%), Linkjapan Eremote (11\%) and JVCKENWOOD Hub (12\%).


\end{document}